# Measurement of the plasma astrophysical S factor for the $^3$He(D, p)$^4$He reaction in exploding molecular clusters

M. Barbui [1, a)], W. Bang [2, b)], A. Bonasera[3,1], K. Hagel[1], K. Schmidt[1], J. B. Natowitz[1], R. Burch[1], G. Giuliani[1], M. Barbarino[1], H. Zheng[1], G. Dyer[2], H. J. Quevedo[2], E. Gaul[2], A. C. Bernstein[2], M. Donovan[2], S. Kimura[4], M. Mazzocco[5], F. Consoli[6], R. De Angelis[6], P. Andreoli[6], T. Ditmire[2]

[1]Cyclotron Institute, Texas A&M University, College Station, TX, 77843, USA

[2]Center for High Energy Density Science, C1510, University of Texas at Austin, Austin, TX, 78712, USA

[3]INFN- Laboratori Nazionali del Sud, via S.Sofia 62, 95125 Catania, Italy

[4]Department of Physics, Università degli Studi di Milano, via Celoria 16, 20133 Milano, Italy

[5]Dipartimento di Fisica e Astronomia Università degli Studi di Padova and INFN Sezione di Padova, Via Marzolo 8, I-35131 Padova, Italy

[6]ENEA,Centro Ricerche Frascati, Via E. Fermi 45, 00044 Frascati (Rome) Italy

The plasma astrophysical S factor for the $^3$He(D, p)$^4$He fusion reaction was measured for the first time at temperatures of few keV, using the interaction of intense ultrafast laser pulses with molecular deuterium clusters mixed with $^3$He atoms. Different proportions of D$_2$ and $^3$He or CD$_4$ and $^3$He were mixed in the gas jet target in order to allow the measurement of the cross-section for the $^3$He(D, p)$^4$He reaction. The yield of 14.7 MeV protons from the $^3$He(D, p)$^4$He reaction was measured in order to extract the astrophysical *S* factor at low energies. Our result is in agreement with other *S* factor parameterizations found in the literature.

Author to whom correspondence should be addressed.  Electronic mail:

a) barbui@comp.tamu.edu

b) dws223@physics.utexas.edu (for experimental details only)

In the past few decades the $^3$He(D, p)$^4$He reaction has been extensively studied from both the experimental and theoretical point of view because of its importance in primordial nucleosynthesis as well as in the design of future nuclear fusion reactors for energy production. Theoretical estimates of this cross-section are obtained from the R-matrix study of the $^5$Li system [1, 2]. Direct [2-4] and indirect [5] measurements of the cross-section have been performed over the years. Direct measurements using accelerated beams show that, at very low energies, the electrons present in the target partially screen the Coulomb barrier between the projectile and the target, resulting in an enhancement of the measured cross-section compared with the bare nucleus cross-section. This screening effect significantly varies with the target composition and prevents a direct measure of the bare nucleus cross-section at the energies of astrophysical interest (1-50 keV) [7]. The indirect measurement of the bare nucleus cross-section presented in Ref. [5] has large error bars at very low energy. The electron screening effect is also present in stellar plasmas and will vary with temperature and density. Therefore, a direct measure of the plasma S factor in the laboratory is very important to have a better understanding of the screening effect in stellar plasmas.

The availability of high intensity laser facilities capable of delivering petawatts of power into small volumes has opened the possibility to use these facilities for fundamental and applied nuclear physics studies. In particular, the Coulomb explosion of $D_2$ molecular clusters induced by the interaction with an intense laser pulse provides a distribution of low energy D ions in a highly ionized medium that might be ideal for the study of $^3$He(D, p)$^4$He nuclear fusion reactions at low energies. In these conditions a direct measurement of the plasma S factor becomes possible [8]. In this work the astrophysical S factor for the $^3$He(D, p)$^4$He fusion reaction was measured for the first time at temperatures of a few keV, using the interaction of intense ultrafast laser pulses with molecular $D_2$ clusters mixed with $^3$He atoms.

The experiment was performed at the Center for High Energy Density Science at The University of Texas at Austin. Laser pulses of energy ranging from 90 to 180 J and 150–270 fs duration were delivered by the Texas Petawatt laser. $D_2$ or $CD_4$ molecular clusters were produced in the adiabatic expansion of a high pressure (52.5 bars) and low temperature gas into vacuum through a supersonic nozzle. The temperature of the gas was 86 K in the case of $D_2$ and 200-260 K in the case of $CD_4$. Different concentrations of $^3$He gas were introduced into the reservoir in order to study collisions between fast D ions and $^3$He. For each laser shot, a residual gas analyzer measured the partial pressures of $D_2$ or $CD_4$ and $^3$He in the reaction chamber. Assuming an isotropic emission of the ions over 4π, the number and the energy distributions of fast D ions were measured with a Faraday cup in a Time-of-Flight (ToF) configuration placed at 107 cm from the interaction volume. The measured ToF spectra show that the energy distributions of the D ions are well described by Maxwellian distributions defined by a temperature $kT_{ToF}$ [9, 10]. The yield of 14.7 MeV protons from the $^3$He(D, p)$^4$He reaction was measured with three thin plastic scintillators placed at 45, 90 and 135 degrees with respect to the laser beam direction. Images of the plasma were taken by different cameras in order to measure the dimensions and calculate the plasma volume. The yield of 2.45 MeV neutrons from the D-D reactions was also measured on each shot and used to obtain an independent measurement of the deuterium ion temperature [11, 12] using D-D and D-$^3$He cross-sections. This temperature was found to be in agreement with the temperature calculated from the ion ToF data. The details of the experimental set-up and the data analysis can be found in Refs. [11, 12].

The cross-section for the $^3$He(D, p)$^4$He reaction can be represented in the Gamow form (energies are in the center of mass system throughout this paper):

$$\sigma(E) = \frac{S(E)}{E} e^{-\sqrt{E_G/E}} \qquad (1)$$

where $S(E)$ is the astrophysical $S$ factor, $E$ is the energy, $E_G$ is the Gamow energy ($E_G$ = 4726.56 keV). This S factor may be parameterized as a function of $E$ in the form of Padé polynomials as used in Ref. [1].

$$S(E) = \frac{a_0 + E(a_1 + a_2 E)}{1 + E(b_0 + E(b_1 + E b_2))} \qquad (2)$$

In Ref. [1], the parameters *a* and *b* were set in order to fit the R-matrix evaluation of the cross-section as a function of the energy. The parameters *b* account for a broad resonance at $E = 210$ keV produced by an excited state of the $^5$Li compound system.

$^3$He atoms do not form clusters in the gas jet expansion process at the temperatures considered in this experiment. They are not accelerated during the Coulomb explosion of deuterium or CD$_4$ clusters and do not efficiently absorb the laser energy. Therefore, D-$^3$He reactions occur between fast D ions and near-stationary $^3$He ions or atoms. Since at each shot the energy distribution of the D ions in the laboratory frame were observed to be characterized by a Maxwellian distribution of temperature $kT_{ToF}$, the energy available in the center of mass frame was a Maxwellian distribution of temperature $kT=3/5kT_{ToF}$. To calculate the yield of 14.7 MeV protons, $Y_p$, we used a simplified model [13] in which $N_D$ deuterium ions are emitted isotropically at the center of the cylindrical plasma, they can travel in the gas jet for a distance $R$ equal to the radius of the gas jet plume at the interaction point (~2.5 mm) and interact with $^3$He. The atomic density per unit area of $^3$He is $\rho_{S,^3He} = \rho_{^3He} R$. In this model, the yield of 14.7 MeV protons produced by D ions of energy $E$ in the interval $dE$ is given by $dY_p = N_D f(E,kT) \sigma(E) \rho_{S,^3He} dE$, where $f(E,kT)$ is a Maxwellian distribution of temperature $kT$ and $\sigma(E)$ is the cross-section from Eq. (2). Therefore, the total proton yield is:

$$Y_p = N_D \rho_{^3He} R \langle \sigma \rangle_{kT}, \tag{3}$$

where

$$\langle \sigma \rangle_{kT} = \int_0^\infty 2\sqrt{\frac{E}{\pi}} \left(\frac{1}{kT}\right)^{3/2} e^{-E/kT} \sigma(E) dE . \tag{4}$$

The atomic density of $^3$He, $\rho_{^3He}$, is obtained as $\rho_{^3He} = \rho_D [^3He]/[D]$, where [$^3$He] and [D] are the concentrations of $^3$He an D in the gas jet. Assuming that all the clusters in the plasma undergo Coulomb explosion, the atomic density of deuterium, $\rho_D$, is given by $N_D$ divided by the measured volume of the plasma. We found an average atomic density, $\rho_D$, of about $2 \times 10^{18}$ atoms/cm$^3$. Energy loss calculations made with the SRIM code [14] suggest that the interaction of the D ions with the surrounding gas may shift down the measured temperature $kT_{ToF}$ by about 5% on average. We applied this correction to the experimental values of $kT_{ToF}$.

From the measured proton yields on each shot, we calculated the experimental values of $<\sigma>_{kT}$, using Eq. (3). The results are shown in Fig. 1 as a function of the temperature, $kT$, and the most effective energy called the Gamow peak energy, $E_{Gp}$. The Gamow peak energy corresponds to the maximum of the function integrated in Eq. (4). At the temperatures considered in this work, $E_{Gp}$ is well described by the value obtained for non-resonant reactions $E_{Gp} \sim (E_G/4)^{1/3} (kT)^{2/3}$ [15]. The points in Fig. 1 are the weighted average values of the experimentally measured $<\sigma>_{kT}$ in $kT$ intervals of 0.3 keV. Among the CD$_4$ shots, we only considered three shots where the ToF spectra showed clear D ion signals. Our experimental points cover the energy range 27 keV $< E_{Gp} <$ 46 keV. At these energies the electron screening effect is very small, as shown in Ref. [4].

A parameterization of the astrophysical factor $S$ was obtained by fitting the points in Fig. 1, using the integral function Eq. (4). Since we do not have any experimental results in the energy region around 200 keV, we fixed the fitting parameters $b_0$, $b_1$, and $b_2$ to the values found in Ref. [1]. The limited energy range covered by our data does not justify the extraction of a second order term, so we fixed $a_2 = 0$ and left only $a_0$ and $a_1$ as free parameters. The solid line in Fig. 1 shows the result of the fit. Using our parameters $a_0 = (5.9 \pm 0.7) \times 10^6$ and $a_1 = (8.0 \pm 1.4) \times 10^3$ we calculated the values of $S(E)$. The results are shown in Fig. 2 where they are compared with the evaluations of the bare nucleus S factor obtained using the indirect Trojan Horse Method (THM) and with those of various direct measurements. Our data agree very well with the THM evaluation of the bare nucleus S factor by La Cognata *et al*. [5]. The agreement with the direct data from Krauss *et al*. [3] is also good. It is interesting to note that the experimental data from Aliotta and Krauss, although obtained in rather similar experimental conditions show a very different screening effect. The data of Geist et al. [2] at energies from 100 keV to 300 keV are fairly reproduced by our parameterization using the parameters *b* from Ref. [1].

This work shows that the interaction of intense ultrafast laser pulses with molecular clusters produces ideal conditions to study the plasma astrophysical *S* factor at low energies. The $^3$He(D, p)$^4$He reactions occurring in exploding $D_2$ ($CD_4$) clusters mixed with $^3$He gas were investigated at Gamow peak energies ranging from 27 keV to 46 keV. From the measurements of the energy distribution of the D ions and the yield of 14.7 MeV protons, we determined the experimental values of $<\sigma>_{kT}$, for *kT* ranging from 4 keV to 10 keV. Fitting these values, we extracted a parameterization for the astrophysical *S* factor, *S(E)*. Our values are in good agreement with the evaluations of the bare nucleus S factor available from Refs. [1, 5], confirming that the electron screening effect is small at the energies explored in this work. This measurement can be extended to energies as low as $E_{Gp}$ = 14 keV (*kT* = 1.5 keV) using smaller deuterium clusters, that are easier to produce, and laser beams of lower intensity but with a higher repetition rate. Irradiating $D_2$ clusters with a 120 mJ, 35 fs laser pulse, Zweiback et al. [6] observed up to about $10^4$ neutrons per shot from the D(D, n)$^3$He reactions and estimated an average D ion energy of about 2.5 keV. Based on our results on $^3$He(D, p)$^4$He fusion reactions, we estimate a proton yield of 56 per shot, assuming a D ion temperature $kT_{ToF}$ = 2.5 keV (*kT* = 1.5 keV), a deuterium density $\rho_D$ = 5×10$^{18}$ atoms/cm$^3$ and a concentration ratio [$^3$He]/[D] = 0.5. With the present experimental setup, this corresponds to one 14.7 MeV proton detected by each proton detector every 200 shots. The interaction of high power laser pulses with molecular clusters can also be used to measure the *S* factor for other systems of astrophysical interest. Low Z systems are preferable in this type of experiment since the penetrability of the Coulomb barrier for high Z systems is extremely small at low energy.

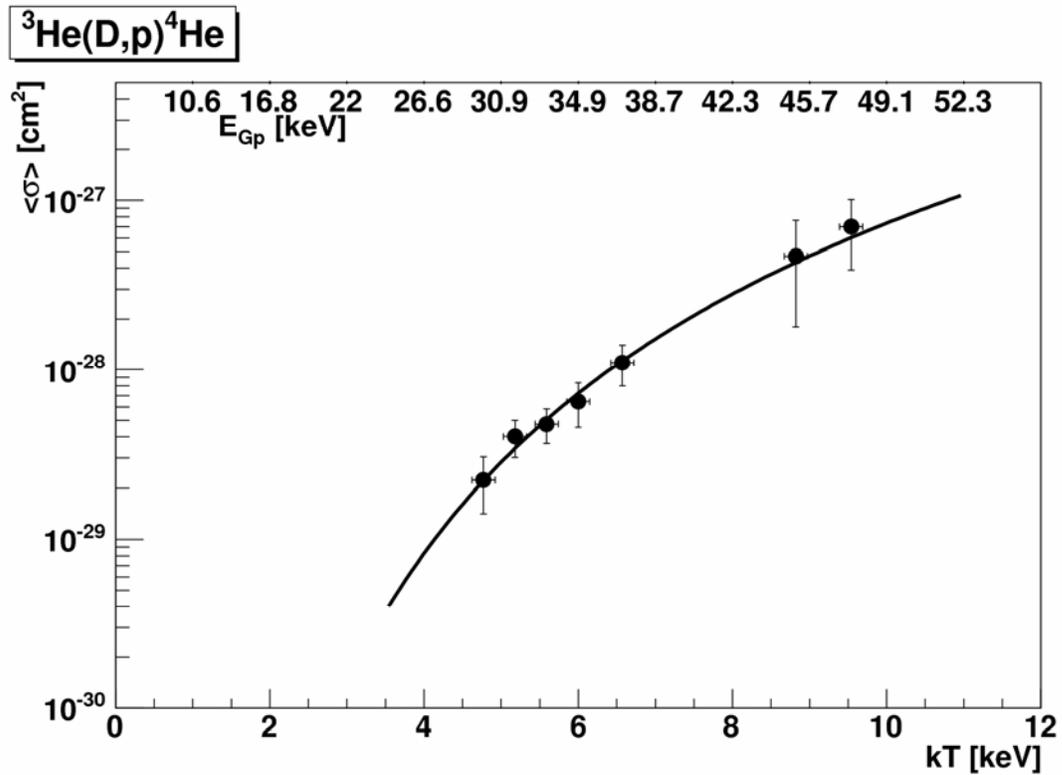

FIG. 1. Experimentally measured $\langle\sigma\rangle_{kT}$ for the $^3$He(D, p)$^4$He reaction as a function of the temperature $kT$. The solid line is the result of the fit with equation (4). See text for details. The upper scale shows the corresponding Gamow peak energy. The $\chi^2$ of the fit is 0.6 with 5 degrees of freedom.

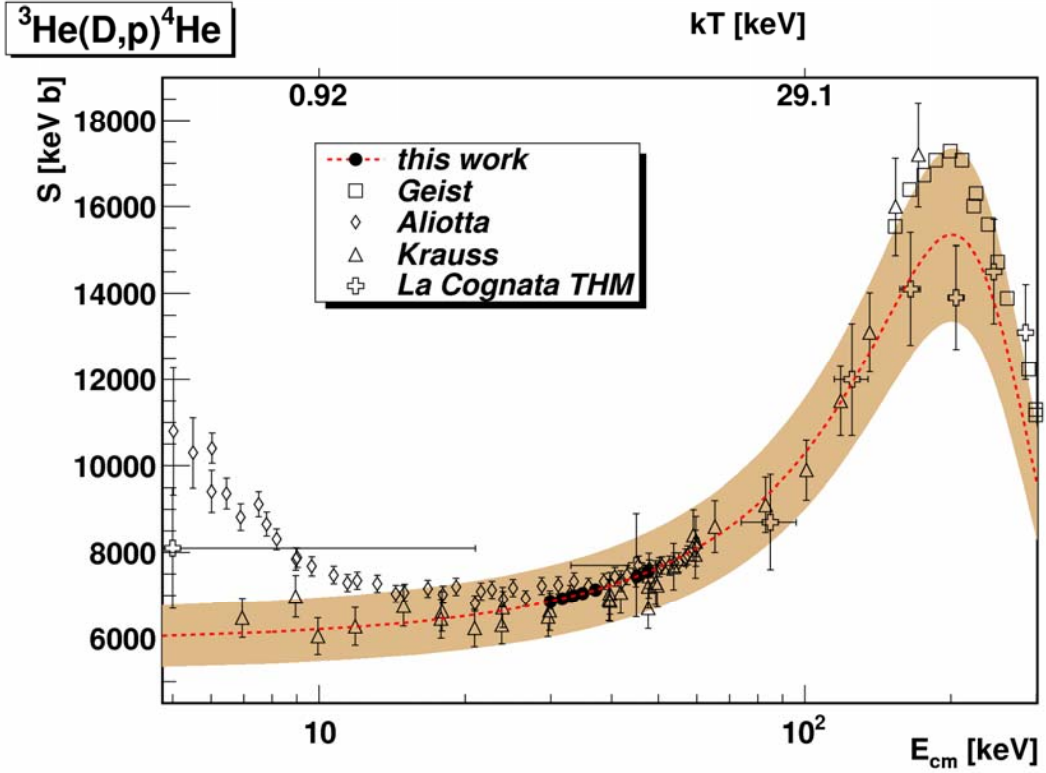

FIG. 2. Astrophysical S-factor for the $^3$He(D, p)$^4$He reaction as a function of the energy in the center of mass frame ($E_{cm}$). The solid black circles show the S factor, calculated from the fit of the experimental $<\sigma>_{kT}$ in Fig.1. The S factor is extrapolated in the energy range 5-300 keV (dashed red line). The shaded area shows the uncertainty on this evaluation. Our result is compared with other experimental data available in Refs. [5] (open crosses), [3] (open triangles), [4] (open diamonds), and [2] (open squares). The upper scale shows the temperature corresponding to the Gamow peak energy in the lower scale.